\documentclass[12pt]{article}

\usepackage{amsmath,amssymb,amsthm}
\usepackage{a4wide}

\newtheorem{theorem}{Theorem}
\newtheorem{lemma}{Lemma}

\title{\bf Dense Dirac combs in Euclidean space\\ 
with pure point diffraction}
\author{\sc Christoph Richard\\
\\
Institut f\"ur Mathematik, Universit\"at Greifswald,\\
Jahnstr.~15a, 17487 Greifswald, Germany}

\date{\today}

\begin{document}

\maketitle

\begin{abstract}
Regular model sets, describing the point positions of ideal
quasicrystallographic tilings, are mathematical models of quasicrystals.
An important result in mathematical diffraction theory of regular model sets,
which are defined on locally compact Abelian groups, is the pure pointedness
of the diffraction spectrum.
We derive an extension of this result, valid for dense point sets in Euclidean
space, which is motivated by the study of quasicrystallographic random
tilings.
\end{abstract}


\section{Introduction}

An important question in mathematical diffraction theory concerns
the problem which distributions of matter diffract 
(Bombieri and Taylor~(1986)).
By now, there are only partial answers to this question which go
beyond the crystallographic case.
A mathematical idealization of the set of atomic positions of a piece
of matter are Delone sets (Lagarias~(2000)).
A subset $\Lambda$ of $\mathbb R^d$ is called a {\it Delone} set if it is
uniformly discrete and relatively dense.
This means that there are radii $r,R>0$ such that each ball of radius 
$r$ (resp. $R$) contains at most (resp. at least) one point of $\Lambda$.
This class, however, is too general to obtain specific results about spectral
properties.
It includes, for example, ordered structures such as crystals, as well as
structures with disorder, and also amorphous systems.

A special class of Delone sets are model sets (see for example
Moody~(1997)),
which arise from a cut-and-project scheme.
We will repeat their precise definition below.
Model sets have strong regularity properties such as uniform point densities.
They are mathematical abstractions of ideal quasicrystals, whose diffraction
spectrum is experimentally known to consist of Bragg peaks only.
By now, the study of model sets is a rather well developing subject, and their
diffraction properties are well understood 
(Hof~(1995), Hof~(1998), Schlottmann~(1998), Schlottmann~(2000)).
The most general description is in terms of measures on locally compact
Abelian groups, in generalization of Euclidean space (Schlottmann~(1998)). 
It is known that regular model sets are pure point diffractive.
A first proof of this fact was given by Hof~(1998)
and, in a more general setup, by Schlottmann~(1998)
, where they used a dynamical 
systems approach together with an argument due to Dworkin~(1993).
Recently, an alternative proof has been given by Baake and Moody~(2002)
, who considered Delone sets with certain additional properties, which
include model sets.
For these sets, they explicitly constructed a cut-and-project scheme
and were able to prove the pure pointedness of the diffraction spectrum
directly.
As they showed, their results have a natural interpretation in terms
of the theory of almost periodic measures (Gil de Lamadrid and
Argabright~(1990)).
An important assumption of the approach is the (uniform) discreteness of
the point set $\Lambda$, arising from the physical motivation.

In this paper, we will show that the diffraction result for regular model sets 
remains valid for certain dense point sets in Euclidean space, 
which appropriately generalize model sets.
These point sets are not a mathematical curiosity but admit an interpretation
in terms of averaged structures derived from {\it random tilings} with
quasicrystallographic symmetries (Henley~(1999), Richard et al.~(1998), Richard~(1999)):
Soon after the discovery of quasicrystals, it became clear that there are
two competing models for the description of their unusual diffraction 
properties, namely the ideal tiling model and the random tiling model.
Whereas the ideal tiling model (i.e., a model set) leads to a diffraction 
spectrum consisting of Bragg peaks only, in the random tiling model a 
continuous background in addition to Bragg peaks was expected, 
as has been argued by Henley~(1999).
There is at present no rigorous treatment of diffraction properties of
quasicrystallographic random tilings in $d>1$, apart from the comparatively
simple situation of disorder arising from independent random variables 
(Baake and Moody~(1998))
and the investigation K\"ulske~(2001).
For rigorous results about diffraction of crystallographic random tilings 
in $d=2$ and random tilings in $d=1$, see Baake and H\"offe~(2000),
H\"offe and Baake~(2000).
The derivation of Henley's prediction rests on non-rigorous arguments and uses
an averaged point distribution, which can be regarded as a generalized model 
set, being dense in Euclidean space.
This distribution was assumed to exist in dimensions $d>2$  (Henley~(1999)).
In the following, we give a rigorous exposition of a main part of this
approach, which is, on the other hand, a generalization of the cut-and-project
scheme for regular model sets to dense point sets (see also H\"offe~(2001),
which introduces the setup used in this paper, and the review article Baake
et al.~(2002),
where the diffraction formula of Theorem \ref{th:ddiff} is cited).
After recollecting the basic terminology and central theorems of diffraction
theory of regular model sets in Section 2, we will define generalized model
sets in Section 3 and show that they are pure point diffractive.
This leads to a generalization of Poisson's summation formula to certain 
dense point sets.
Section 4 will apply the results within Henley's framework for the description
of diffraction properties of quasicrystallographic random tilings. 
A discussion of open questions and possible future work concludes the paper.

\section{Diffraction of weighted regular model sets}

In this section, we recollect and extend results from diffraction
theory of weighted regular model sets, which we shall generalism in
the following section to the setup of a dense Dirac comb.

Throughout the paper, we consider the situation of a 
{\it cut-and-project scheme} (Moody (1997)) 
with a Euclidean space $\mathbb R^d$ and a Euclidean space $\mathbb
R^m$, called {\it direct} space and {\it internal} space:

Let ${\widetilde L}\subset\mathbb R^d \times \mathbb R^m$ be a lattice
and denote the volume of its fundamental domain by $|\mbox{\rm
det}({\widetilde L})|$. 
If $\pi_1$ and $\pi_2$ are the natural (orthogonal) projections of 
$\mathbb R^d\times\mathbb R^m$ onto $\mathbb R^d$ and $\mathbb R^m$,
respectively, then $\pi_1|_{\widetilde L}$ is assumed to be one-to-one
and $\pi_2({\widetilde L})$ is assumed to be dense.
Set $L=\pi_1(\widetilde L)$ and let $()^\star:L\to\pi_2(\widetilde L)$ 
denote the mapping $\pi_2\circ (\left.\pi_1\right|_{\widetilde L})^{-1}$, 
also called the {\it star-map}.
This is summarized in the following diagram.

\begin{equation}
\renewcommand{\arraystretch}{1.5}
\begin{array}{ccccc}
& \pi_1 & & \pi_2 & \vspace*{-1.5ex} \\
\mathbb R^{d} & \longleftarrow & \mathbb R^{d}\times \mathbb R^m &
\longrightarrow & \mathbb R^m \\ 
\cup  & \mbox{\raisebox{-1.5ex}{\footnotesize
     \textnormal{1--1}}}\!\!\!\!\nwarrow\;\; & \cup &  
\;\;\nearrow\!\!\!\!\mbox{\raisebox{-1.5ex}{\footnotesize \textnormal{dense}}}
& \\
L & & \tilde{L}=\{(x,x^\star)\, | \, x\in L\} & &
\end{array}
\end{equation} 
For an open and relatively compact subset $W\subset\mathbb R^m$, also
called {\it window}, define the {\it model set} $\Lambda(W)$ by
\begin{equation}
\Lambda(W) := \{x\in L\, |\, x^\star \in W\}.
\end{equation}
Denote the volume of $W$ by $\mbox{vol}(W)$.
Model sets are Delone sets, i.e., they are both uniformly discrete and
relatively dense (Moody~(1997), Schlottmann~(1997)).
The model set $\Lambda(W)$ is called {\it regular} if $\emptyset\neq W =
\overline{\mbox{int}(W)}$ is compact and if $\partial W$ has zero Lebesgue
measure.

Associated with the lattice $\widetilde L$ is its dual lattice $(\widetilde
L)^*$, defined via 
\begin{equation}
(\widetilde L)^*=\{y\in \mathbb R^d\times\mathbb R^m\,|\,
y\cdot x\in \mathbb Z \mbox{ for all } x\in\widetilde L\}.
\end{equation}
Denote its image in $\mathbb R^d$ by $L^*=\pi_1((\widetilde L)^*)$.
A remarkable property of cut-and-project schemes is duality (Moody~(1997), p.~418).
We can dualize the given cut-and-project scheme to obtain a cut-and-project 
scheme for the dual lattice.
By identifying direct and internal space with their corresponding duals, 
we can see that the projection $\pi_1$, restricted to the dual lattice
$(\widetilde L)^*$, is one-to-one, and that the dual lattice has dense image
$\pi_2((\widetilde L)^*)$ in $\mathbb R^m$.
The corresponding star-map is defined on the set $L^*$.
We denote it by $()^\star$ again.

Regular model sets have a well-defined density.
Throughout the paper, we will consider infinite volume limits to be taken on
sequences of balls.
Existence of these limits may however be derived for more general van Hove
sequences, see Schlottmann~(2000)
for the example of the density formula.
Let $B_n(a)$ denote the closed ball of radius $n$ centered at
$a\in\mathbb R^d$.  We set $B_n=B_n(0)$.

\begin{theorem}[Density formula {\rm (Schlottmann (1998))}]
Let $\Lambda(W)$ be a regular model set.
Then
\begin{displaymath}
\lim_{r \to \infty} \frac{1}{\mbox{\rm vol}(B_r(0))} 
\left( \sum_{x\in\Lambda(W+u) \cap B_r(a)} 1 \right)
= \frac{\mbox{\rm vol}(W)}{|\det ({\widetilde L})|},
\end{displaymath}
uniformly in $a$ and in $u$.\qed
\end{theorem}

This result may be used to consider sums over weighted regular model sets.
Consider first the situation where the weight of a point $x\in\Lambda(W)$ only
depends on $x$.

\begin{lemma}\label{modsum}
Let $\Lambda(W)$ be a regular model set, $f:\mathbb R^d \to \mathbb C$ bounded 
and $|x|^{d+1+\alpha}|f(x)|\le C$ for some constants $C>0$ and $\alpha>0$.
Then, for $a\in\mathbb R^d$ and $u\in\mathbb R^m$, the sums 
\begin{displaymath}
s(u, a):= \sum_{x\in \Lambda(W+u)+a} f(x)
\end{displaymath}
are absolutely convergent.
Moreover, they are uniformly bounded in $u$ and in $a$, the bound being
proportional to $\mbox{\rm vol}(W)$.
\end{lemma}

\begin{proof}
For $R\in\mathbb N$, consider the sums
\begin{displaymath}
s_<(u,a,R) := \sum_{\genfrac{}{}{0pt}{1}{x\in \Lambda(W+u)+a}{|x| < R}}
|f(x)|, 
\qquad 
s_\ge(u,a,R) := \sum_{\genfrac{}{}{0pt}{1}{x\in \Lambda(W+u)+a}{|x| \ge R}}
|f(x)|.
\end{displaymath}
In the density formula, the sequence on the lhs is certainly bounded
by twice its limit for almost all $r$.
This implies for the number of points within a ball of radius $n$
\begin{displaymath}
|\left( \Lambda(W+u)+a\right) \cap B_n(0)|=|\Lambda(W+u)\cap B_n(-a)| \le 
c \, \mbox{vol}(W) \, n^d \quad (n>n_0(W)),
\end{displaymath}
uniformly in $u$ and in $a$, where $c=2S_d/|\det ({\widetilde L})|$,
and $S_d \subset \mathbb R^d$ denotes the volume of the unit ball.
The number $n_0(W)$ is increasing with decreasing volume of $W$.
Together with $f$ bounded, the above estimate implies that $s_<(u,a,R)$ is
uniformly bounded in $u$ and in $a$ with a number proportional to $\mbox{\rm
  vol}(W)$.
We derive a uniform bound on $s_\ge(u,a,R)$.
By assumption, we have $|x|^{d+1+\alpha} |f(x)| \le C$.
Define $g_n := \sum_{n\le |x| < n+1} |f(x)|$ and estimate $g_n$ by
\begin{equation}\label{form:est2}
g_n \le |\left(\Lambda(W+u)+a\right)\cap B_{n+1}(0)| \,
\frac{C}{n^{d+1+\alpha}}
\le c\,\mbox{vol}(W)(n+1)^d \frac{C}{n^{d+1+\alpha}}
\le \frac{2c\,C\,\mbox{vol}(W)}{n^{1+\alpha}}
\end{equation}
for $n>n_1(W)$.
Thus for $R>n_1(W)$
\begin{equation}\label{form:est}
|s_\ge(u,a, R)| \le  \sum_{n \ge R} g_n \le 
2c\,C\,\mbox{vol}(W) \sum_{n=R}^\infty \frac{1}{n^{1+\alpha}} < \infty, 
\end{equation}
since by assumption $\alpha>0$.
The bound is independent of $u$ and of $a$. 
Note that we have $\lim_{R \to \infty} s_\ge(u,a,R)=0$.
\end{proof}

\noindent {\bf Remark.}
The above sums $s(u,a)$ are absolutely convergent and bounded uniformly in $u$
and in $a$ under the milder assumption $|x|^{d+\alpha}|f(x)|\le C$ for
some constants $C>0$ and $\alpha>0$.
This follows already from the uniform discreteness of $\Lambda(W)$ by a
standard estimation.
The additional property that there exists a bound proportional to $\mbox{\rm
  vol}(W)$, which we will use extensively below, is a result of the density
formula.
It would be interesting to consider whether the assumptions on $f$ in the above
lemma may be weakened, since the estimation in equation (\ref{form:est2})
seems rather crude.
\vspace{1ex}

Consider now the situation where the weight of a point $x\in\Lambda(W)$
depends only on its internal coordinate $x^\star$.
This leads to Weyl's theorem on uniform distribution in the context of regular
model sets (Kuipers and Niederreiter~(1974), Schlottmann~(1998)).

\begin{theorem}[Weyl's Theorem for regular model sets {\rm (Baake and
Moody~(2000))}] 
Let $\Lambda(W)$ be a regular model set, with compact, Riemann measurable $W
\subset \mathbb R^m$.
Let $f : \mathbb R^m \to \mathbb C$ be continuous with $\mbox{\rm supp}(f)
\subset W$.
Then, for all $a\in\mathbb R^d$,
\begin{displaymath}
\lim_{r \to \infty} \frac{1}{\mbox{\rm vol}(B_r(0))} 
\sum_{x\in\Lambda(W) \cap B_r(a)} f(x^\star) 
= \frac{1}{|\det ({\widetilde L})|} \int_{W} f(y) \, dy,
\end{displaymath}
uniformly in $a$.\qed
\end{theorem}

The fundamental object in diffraction theory of weighted model sets is the
{\it weighted Dirac comb} $\omega$ defined by
\begin{equation}
\omega = \sum_{x\in\Lambda(W)} f(x^\star) \delta_x,
\end{equation}
where $\mbox{sup}_{x\in\Lambda(W)} |f(x^\star)|<\infty$.
This defines a complex regular Borel measure on $\mathbb R^d$, 
which is translation bounded, since $\Lambda(W)$, being a model set, 
is uniformly discrete.
Recall that a measure $\omega$ is {\it translation bounded} iff, for all
compact $K\subset\mathbb R^d$, $\mbox{sup}_{y\in\mathbb R^d} |\omega|(y+K)\le C_K<\infty$
for some constant $C_K$ which only depends on $K$.
Here, $|\omega|$ denotes the total variation measure and $y+K=\{y+x\,|\, x\in
K\}$.
In the following, we assume that the function $f$ satisfies the assumption of
Weyl's theorem for regular model sets, i.e., we assume that $f : \mathbb R^m \to \mathbb C$ is continuous
with $\mbox{\rm supp}(f) \subset W$.

Diffraction properties can be expressed using the {\it Fourier-Bohr
  coefficient} $c_W(k)$ of the weighted Dirac comb $\omega$, defined by
\begin{equation}
c_W(k) = \lim_{r\to\infty} \frac{1}{\mbox{\rm vol}(B_r(0))} 
\sum_{x\in\Lambda(W)\cap B_r(a)} f(x^\star) e^{-2\pi \imath k\cdot x},
\end{equation}
where $k\in\mathbb R^d$ and $a\in\mathbb R^d$.
We have the following theorem.

\begin{theorem}[Fourier-Bohr coefficients {\rm (Bernuau and Duneau
(2000), Hof (1995))}]
\label{th:FBreg}
Let $\Lambda(W)$ be a regular model set, with compact, Riemann measurable $W
\subset \mathbb R^m$.
Let $f : \mathbb R^m \to \mathbb C$ be continuous with $\mbox{\rm supp}(f)
\subset W$.
Then, for all $k\in\mathbb R^d$, the Fourier-Bohr coefficient $c_W(k)$ exists
and is independent of $a\in\mathbb R^d$.
Its value is given as follows.
For any $k\in L^*$, one has
\begin{displaymath}
c_W(k) = \frac{1}{|\det ({\widetilde L})|} \int_W e^{2\pi \imath
  k^\star\cdot u} f(u) \, du
= \frac{1}{|\det ({\widetilde L})|} {\widehat {f|_W}}(-k^\star),
\end{displaymath}
and $c_W(k)=0$ if $k$ does not belong to the $\mathbb Z$-module $L^*$.\qed
\end{theorem}

\noindent {\bf Remark.}
In Bernuau and Duneau~(2000), 
the theorem is proved only for the case where $f(u)$ equals unity.
The statement can be generalized to the situation described above by the same
methods which lead to Weyl's formula, generalizing the density formula.
For $k\in L^*$, the theorem is a direct consequence of Weyl's density formula.
\vspace{1ex}

For regular model sets, the {\it weighted density of points} $\rho$ exists,
because Weyl's theorem for regular model sets implies
\begin{equation}
\rho := \lim_{r\to\infty} \frac{1}{\mbox{vol}(B_r)}\, \omega(B_r)
=\frac{1}{|\det ({\widetilde L})|} \int_W f(u) \, du.
\end{equation}
This identity may be viewed as a particular normalization of admissible
functions $f$, which we employ in the following.

Diffraction is described by properties of the Fourier transform of the 
{\it autocorrelation}, which we now define (see also Hof~(1995),
Baake, Moody and Pleasants~(2000)).
Set $\widetilde{\omega}(f)=\overline{\omega(\tilde f)}$, 
where $\tilde f(x)=\overline{f(-x)}$.
Define truncated Dirac combs $\omega_n=\left. \omega \right|_{B_n}$ and set
${\widetilde \omega}_n=(\omega_n)\tilde{}$.
The finite autocorrelation measures
\begin{equation}
\gamma_\omega^{(n)} := \frac{1}{\mbox{vol}(B_n)}\,\omega_n *
{\widetilde\omega}_n
\end{equation}
are well defined, since $\omega_n$ has compact support.
Recall that the {\it convolution} of two measures $\mu,\nu$ is defined as
$\mu*\nu(f)=\int_{\mathbb R^d\times\mathbb R^d}f(x+y)\,d\mu(x)\,d\nu(y)$, being
well-defined if at least one of them has compact support. 
The finite autocorrelation measures read explicitly
\begin{equation}
\gamma_\omega^{(n)}=\sum_{z\in \Delta}\eta_n(z)\delta_z, \qquad
\eta_n (z) = \frac{1}{\mbox{vol}(B_n)} 
\sum_{\genfrac{}{}{0pt}{1}{x,y\in \Lambda(W) \cap B_n}{x-y=z}} f(x^\star)
\overline{f(y^\star)},
\end{equation}
where $\Delta=\Lambda(W)-\Lambda(W)$ is the set of 
difference vectors of $\Lambda(W)$.
It can be shown that the vague limit $n\to\infty$ leads to a unique 
autocorrelation $\gamma_\omega$.

\begin{theorem}[Autocorrelation]\label{form:autoc}
Let $\Lambda(W)$ be a regular model set, with compact, 
Riemann measurable $W \subset \mathbb R^m$.
Let $f : \mathbb R^m \to \mathbb C$ be continuous with $\mbox{\rm supp}(f)
\subset W$.
Then, the vague limit $n\to\infty$ of the finite autocorrelation measures
$\gamma_\omega^{(n)}$ leads to a unique measure $\gamma_\omega$,
called natural autocorrelation, being a translation bounded,
positive definite pure point measure.
It is given explicitly by
\begin{equation}
\gamma_\omega=\sum_{z\in \Delta}\eta(z)\delta_z, \qquad
\eta(z) = \frac{1}{|\det ({\widetilde L})|} 
\int_{W\cap (W+z^\star)} f(u) \overline{f(u-z^\star)}\, du.
\end{equation}\qed
\end{theorem}

Recall that a measure $\mu$ is {\it positive definite} iff
$\mu(g*{\widetilde g})\ge0$ for all compactly supported continuous 
functions $g$.

The proof of the theorem proceeds as follows (Baake, Moody and
Pleasants~(2000)).
All autocorrelation coefficients $\eta(z)=\lim_{n\to\infty}\eta_n(z)$ exist
due to Weyl's theorem for regular model sets.
They are locally summable, since $\Delta$ is closed and discrete.
Then, $\gamma_\omega$ defines a distribution over the space of all
$C^\infty$-functions of compact support.
The translation boundedness is inherited from $\omega$ (see Hof~(1995), Prop.~2.2).
Finally, $\gamma_\omega$ can be written as a certain volume-normalized
convolution, which implies that it is a distribution of positive type.

Due to Bochner's theorem (Reed and Simon~(1980), p.~331),
the Fourier transform $\widehat{\gamma_\omega}$ of $\gamma_\omega$ 
is a positive measure and also translation bounded (Baake, Moody and
Pleasants (2000)).
Recall that the {\it Fourier transform} $\widehat \mu$ of a tempered 
distribution $\mu$ is defined as $\widehat \mu(\varphi)=\mu(\widehat \varphi)$ 
for all Schwartz functions $\varphi$, where
$\widehat \varphi(y)=\int e^{-2\pi\imath y\cdot x}\varphi(x)\, dx$.
It can be shown that the Fourier transform of the autocorrelation
is a pure point measure (Schlottmann (2000), Baake and Moody (2002)).

\begin{theorem}[Diffraction formula {\rm (Bernuau and Duneau
(2000))}]
\label{th:diff}
Let $\Lambda(W)$ be a regular model set, with compact, Riemann measurable
$W\subset \mathbb R^m$.
Let $f : \mathbb R^m \to \mathbb C$ be continuous with $\mbox{\rm supp}(f)
\subset W$.
The Fourier transform $\widehat{\gamma_{\omega}}$ of the autocorrelation 
measure $\gamma_\omega$ is a translation bounded, positive pure point measure. 
It is explicitly given by
\begin{displaymath}
\widehat{\gamma_{\omega}} = \sum_{k\in L^*} |c_W(k)|^2 \delta_k,
\end{displaymath}
where $L^*$ is the projection of the dual lattice into direct space,
and $c_W(k)$ are the Fourier-Bohr coefficients of Theorem \ref{th:FBreg}.\qed
\end{theorem}

Recall that a measure $\mu$ is {\it positive} iff $\mu(g)\ge0$ for all
compactly supported continuous functions $g\ge0$.
\vspace{2ex}

\noindent {\bf Remark.}
The pure pointedness of $\widehat{\gamma_{\omega}}$ has been shown
in Hof (1998), Schlottmann (1998), Baake and Moody (2002).
The explicit formula for the discrete part of $\widehat{\gamma_{\omega}}$
was proved in Hof (1995),
but appeared earlier at different places in the physical literature.
In the context of deformed model sets, which include regular model sets 
as a special case, the theorem appears in (Bernuau and Duneau
(2000)).
The diffraction formula is often stated for unweighted Dirac combs, but also
holds for weighted Dirac combs, as is seen by an approximation of the weight
function $f$ by step functions, analogously to the proof of Weyl's
theorem for regular model sets in Baake and Moody (2000)
using the density formula.
\vspace{1ex}

\section{Diffraction of dense Dirac combs}

We now extend the above results to the situation of a dense Dirac comb.
The corresponding proofs will rely on the above results.
In the following, we assume a cut-and-project scheme as in the previous
section with the additional property that the canonical projections $\pi_1$
and $\pi_2$, restricted to $\widetilde L$, are both one-to-one, and the images
of $\widetilde L$ are both dense in $\mathbb R^d$ and $\mathbb R^m$,
respectively.
The star-map is then a bijection between $L$ and $L^\star$.
This additional assumption makes it possible to regard subsets
$B\subset\mathbb R^d$ as windows, leading to model sets in internal
space $\mathbb R^m$.

Define the {\it weighted Dirac comb} $\omega$ by
\begin{equation}
\omega = \sum_{x\in L} f(x^\star) \delta_x.
\end{equation}
Since the $\mathbb Z$-module $L$ is dense in Euclidean space,
the above sum is now well defined only under special assumptions
on the weight function $f$.

\begin{theorem}[Weighted Dirac combs with dense support]\label{comb}
Let $f:\mathbb R^m \to \mathbb C$ be bounded, and
$|x|^{m+1+\alpha}|f(x)|\le C$ for some constants $C>0$ and $\alpha>0$.
Then, the weighted Dirac comb $\omega$ is a translation bounded measure.
\end{theorem}

\begin{proof}
Let a compact $K \subset \mathbb R^d$ be given.
Cover $K$ with a finite number of translated unit balls $W_i$ such that 
$K \subset \bigcup_{i=1}^n W_i = W$.
For $y\in\mathbb R^d$, the total variation measure of $\omega$ is bounded by
\begin{displaymath}
|\omega|(K+y)\le \sum_{x \in L\cap (W+y)} |f(x^\star)| = \sum_{x^\star \in
 \left(L\cap (W+y) \right)^\star} |f(x^\star)| =: a_K(y).
\end{displaymath}
The additional assumptions on the cut-and-project setup, introduced at
the beginning of this section, imply that $\left( L\cap (W+y)
\right)^\star$ is a regular model set with window $(W+y)\subset\mathbb R^d$.
Lemma \ref{modsum} then yields that $a_K(y) < b_K < \infty$,
uniformly in $y$, for some $b_K$.
This implies that $\omega$ is a measure, and that $\omega$ is translation
bounded.
\end{proof}

We derive a generalization of Weyl's theorem on uniform distribution
to dense point sets.
\begin{theorem}[Weyl's Theorem for dense point sets]\label{Weyl2}
Let $f:\mathbb R^m \to \mathbb C$ be continuous, and
$|x|^{d+1+\alpha}|f(x)|\le C$ for some constants $C>0$ and $\alpha>0$.
Then
\begin{displaymath}
\lim_{r \to \infty} \frac{1}{\mbox{\rm vol}(B_r(0))} 
\sum_{x\in L \cap B_r(a)}
f(x^\star) = \frac{1}{|\det ({\widetilde L})|} \int_{\mathbb R^m}
f(u) \, du,
\end{displaymath}
uniformly in $a$.
\end{theorem}

\begin{proof}
For $s\in\mathbb N$, let $B_s\subset\mathbb R^m$ denote the ball of
radius $s$ centered at $0$.
The idea of the proof is to approximate the dense set $L$ by the
model sets $\Lambda(B_s)$.
Let $\chi_s:\mathbb R^m\to[0,1]$ be continuous with $\chi_s(x)=1$ for
$|x|<s-1$ and $\chi_s(x)=0$ for $|x|\ge s$.
Define the numbers
\begin{displaymath}
\begin{split}
w_{r,s}  = \frac{1}{\mbox{vol}(B_r(0))} \sum_{x\in \Lambda(B_s) \cap B_r(a)}
(\chi_s\cdot f)(x^\star),
&\qquad
w'_s = \frac{1}{|\det ({\widetilde L})|} \int_{B_s} (\chi_s\cdot
f)(u) \, du,\\
w_r  = \frac{1}{\mbox{vol}(B_r(0))} \sum_{x\in L \cap B_r(a)} f(x^\star),
&\qquad
w = \frac{1}{|\det ({\widetilde L})|} \int_{\mathbb R^m} f(u) \, du.
\end{split}
\end{displaymath}
We have $\lim_{r\to\infty} w_{r,s}=w'_s<\infty$ due to Weyl's theorem
for regular model sets.
We also have $\lim_{s \to \infty} w'_s = w < \infty$.
This follows from Lebesgue's dominated convergence theorem, since
the functions $|(\chi_s\cdot f)(u)|$ are bounded by $|f(u)|$ uniformly
in $s$, and $f(u)$ is by assumption absolutely integrable 
(the integral being finite).

We now show that $\lim_{s \to \infty} w_{r,s} = w_r$ uniformly in $r$ and in
$a$.
We have the estimate
\begin{eqnarray*}
|w_r-w_{r,s}|
&=&
\frac{1}{ \mbox{vol}(B_r(0))}
\left|
\sum_{x\in L\cap B_r(a)} f(x^\star)-\sum_{x\in \Lambda(B_s)\cap B_r(a)}
(\chi_s\cdot f)(x^\star)
\right|
\\ 
&\le &
\frac{1}{ \mbox{vol}(B_r(0))}
\left(
\sum_{x^\star\in (L\cap B_r(a))^\star \setminus B_s} |f(x^\star)| +
\sum_{\genfrac{}{}{0pt}{1}{x^\star\in (L \cap
    B_r(a))^\star}{s-1\le|x^\star|<s}} |f(x^\star)|
\right)
\\
&=&
\frac{1}{ \mbox{vol}(B_r(0))}
\left(
\sum_{x^\star\in (L\cap B_r(a))^\star \setminus B_{s-1}} |f(x^\star)|
\right).
\end{eqnarray*}
We estimate the last term using equation (\ref{form:est}) in Lemma
\ref{modsum}.
Since $n_1(B_{r_2}) \le n_1(B_{r_1})$ for $r_2\ge r_1$,
we get for $s \ge n_1(B_1)$ and $r>1$
\begin{displaymath}
|w_r-w_{r,s}| \le 
\frac{1}{ \mbox{vol}(B_r)} 
\left(
2c\,C\,\mbox{vol}(B_r) \sum_{n=s-1}^\infty \frac{1}{n^{1+\alpha}}
\right).
\end{displaymath}
Since the bound is independent of $r$ and $a$ and vanishes as $s\to\infty$, the
assertion follows.

We now use a $3\varepsilon$ argument to show that $w_r\to w$.
Fix $\varepsilon>0$.
We have $| w - w'_s|<\varepsilon$ for 
$s>s_0(\varepsilon)$.
We also have $|w_{r,s} -w_r|<\varepsilon$ for 
$s>s_1(\varepsilon)$ uniformly in $r$.
Lastly, we have $|w'_s-w_{r,s}|<\varepsilon$ for 
$r>r_0(\varepsilon,s)$.
Take $s>\max \{s_0(\varepsilon), s_1(\varepsilon) \}$.
For $r>r_0(\varepsilon,s)$, we thus have
\begin{displaymath}
|w-w_r| \le  | w - w'_s| + |w'_s-w_{r,s}| + |w_{r,s}-w_r|\le 3\,\varepsilon.
\end{displaymath}
This establishes the convergence result of the theorem.
\end{proof}

For applications to diffraction, we restrict the class of admissible functions
$f$ in the remainder of this paper.
Let $f:\mathbb R^m \to \mathbb C$ be continuous,
and $|f(x)|\le C/|x|^{m+1+\alpha}$ for some constants $C>0$ and $\alpha>0$.
We can argue as in the previous section, using Weyl's theorem for 
dense point sets, that the weighted density of points exists,
\begin{equation}
\rho := \lim_{r\to\infty} \frac{1}{\mbox{vol}(B_r)}\, \omega(B_r)
=\frac{1}{|\det ({\widetilde L})|} \int_{\mathbb R^m} f(u) \, du.
\end{equation}
This identity may be viewed as a particular normalization of 
admissible functions $f$.
Given a weighted Dirac comb $\omega$ with dense support, 
we consider its Fourier-Bohr coefficients.
We define the finite volume approximations
\begin{equation}
c_r(k) =  \frac{1}{\mbox{\rm vol}(B_r(0))} 
\sum_{x\in L \cap B_r(a)} f(x^\star) e^{-2\pi\imath k\cdot x},
\end{equation}
where $k\in\mathbb R^d$ and $a\in\mathbb R^d$.
The numbers $c_r(k)$ exist (and may depend on $a$), as is seen by an argument
similar to that used for proving the existence of the Dirac comb $\omega$.
In the limit $r\to\infty$, we have the following result.

\begin{theorem}[Fourier-Bohr coefficients for dense point sets]\label{th:dFB}
Let $f:\mathbb R^m \to \mathbb C$ be continuous, and 
$|x|^{m+1+\alpha}|f(x)|\le C$ for some constants $C>0$ and $\alpha>0$.
Then, for all $k\in\mathbb R^d$, the Fourier-Bohr coefficient
$c(k)=\lim_{r\to\infty} c_r(k)$ exists and is independent of $a$.
Its value is given as follows.
For any $k\in L^*$ and for any $a\in\mathbb R^d$, one has
\begin{displaymath}
c(k) = \frac{1}{|\det ({\widetilde L})|} \int_{\mathbb R^m} e^{2\pi
  \imath k^\star\cdot u} f(u) \, du
= \frac{1}{|\det ({\widetilde L})|} {\widehat f}(-k^\star),
\end{displaymath}
and $c(k)=0$ if $k$ does not belong to the $\mathbb Z$-module $L^*$.
\end{theorem}

\begin{proof}
The proof is analogous to the proof of Weyl's theorem for dense point sets
above.
For $s\in\mathbb N$, let $B_s\subset\mathbb R^m$ denote the ball of
radius $s$, centered at $0$, and define the numbers
\begin{displaymath}
\begin{split}
&c_{r,s}(k)  = \frac{1}{\mbox{vol}(B_r(0))} \sum_{x\in \Lambda(B_s) \cap
  B_r(a)} (\chi_s\cdot f)(x^\star) e^{-2\pi\imath k\cdot x},\\
&c'_s(k) = \left\{ 
\begin{array}{cc}
\frac{1}{|\det ({\widetilde L})|} \int_{B_s} (\chi_s\cdot f)(u)
e^{2\pi\imath k^\star\cdot u} du,
& k\in L^*\\
0, & k\notin L^*
\end{array}
\right.,\\
&c'(k) = \left\{ 
\begin{array}{cc}
\frac{1}{|\det ({\widetilde L})|} \int_{\mathbb R^m} f(u)
e^{2\pi\imath k^\star\cdot u} du,
& k\in L^*\\
0, & k\notin L^*
\end{array}
\right..
\end{split}
\end{displaymath}
We have $\lim_{r\to\infty} c_{r,s}(k)=c'_s(k)<\infty$ due to Theorem
\ref{th:FBreg}.
We also have $\lim_{s \to \infty} c'_s (k)= c'(k) < \infty$ by Lebesgue's
dominated convergence theorem.
We show that $\lim_{s \to \infty} c_{r,s}(k) = c_r(k)$ uniformly in $r$ and in
$a$.
As in Theorem \ref{Weyl2}, we can then use a $3\varepsilon$ argument to show
that $c_r(k)$ converges to $c'(k)$ as $r\to\infty$.  Thus
$c(k)=\lim_{r\to\infty} c_r(k)=c'(k)$, and the assertion of the
theorem follows.

For every $k\in\mathbb R^d$, we have the estimate
\begin{eqnarray*}
|c_r(k)-c_{r,s}(k)| 
&\le &
\frac{1}{ \mbox{vol}(B_r(0))}
\left(
\sum_{x^\star\in (L\cap B_r(a))^\star \setminus B_s} |f(x^\star)| +
\sum_{\genfrac{}{}{0pt}{1}{x^\star\in (L \cap
    B_r(a))^\star}{s-1\le|x^\star|<s}} |f(x^\star)|
\right).
\end{eqnarray*}
This is the same expression as in the proof of Theorem \ref{Weyl2}.
We can repeat the previous argument and derive a uniform bound in $r$ and in
$a$, which vanishes as $s\to\infty$.
\end{proof}

The finite autocorrelation measures of the dense Dirac comb $\omega$,
\begin{equation}
\gamma_\omega^{(n)} := \frac{1}{\mbox{vol}(B_n)}\,\omega_n *
{\widetilde\omega}_n,
\end{equation}
are well defined, since $\omega_n$ has compact support.
They read explicitly
\begin{equation}
\gamma_\omega^{(n)}=\sum_{z\in L}\eta_n(z)\delta_z, \qquad
\eta_n (z) = \frac{1}{\mbox{vol}(B_n)} 
\sum_{\genfrac{}{}{0pt}{1}{x,y\in L \cap B_n}{x-y=z}} f(x^\star)
\overline{f(y^\star)}.
\end{equation}
We show that the limit $n\to\infty$ leads to a unique autocorrelation.
As a first step, we show that the pointwise limit exists.

\begin{lemma}\label{form:faut}
Let $\eta_n(z)$ denote the finite autocorrelation coefficients of the weighted
Dirac comb $\omega$.
The numbers $\eta_n(z)$ have a well-defined limit $\eta(z)$, which is a
positive definite function.
It is given by
\begin{equation}
\lim_{n \to \infty} \eta_n(z) = \eta(z)= 
\frac{1}{|\det ({\widetilde L})|} \int_{\mathbb R^m} f(u)
\overline{f(u-z^\star)} \, du.
\end{equation}
\end{lemma}

\begin{proof}
This is an application of Weyl's theorem for dense point sets.
We have
\begin{eqnarray*}
\eta_n(z) &=& \frac{1}{\mbox{vol}(B_n)} 
\sum_{\genfrac{}{}{0pt}{1}{x\in L \cap B_n}{x-z\in L \cap B_n}} f(x^\star)
\overline{f(x^\star-z^\star)}\\
&=&\frac{1}{\mbox{vol}(B_n)}
\sum_{x\in L \cap B_n} f(x^\star) \overline{f(x^\star-z^\star)} -
\frac{1}{\mbox{vol}(B_n)} 
\sum_{\genfrac{}{}{0pt}{1}{x\in L \cap B_n}{x-z\notin L \cap B_n}} f(x^\star)
\overline{f(x^\star-z^\star)}.
\end{eqnarray*}
The first term in the last line converges to $\eta(z)$ by Weyl's
theorem for dense point sets, and the second one, which we denote by
$r_n(z)$, converges to zero, as we now show.
Note that $|x|<n-|z|$ implies $|x-z|<n$.
Consider for $n>|z|$ the estimate
\begin{displaymath}
\begin{split}
|r_n(z)| 
&\le 
\frac{1}{\mbox{vol}(B_n)} 
\left( \sum_{x\in L \cap B_n} - \sum_{x\in L \cap B_{n-|z|}} \right)
|f(x^\star)|\, |f(x^\star-z^\star)|\\
&=
\frac{1}{\mbox{vol}(B_n)} \sum_{x^\star\in (L \cap (B_n\setminus
  B_{n-|z|}))^\star} |f(x^\star)|\, |f(x^\star-z^\star)|.
\end{split}
\end{displaymath}
This is a summation over a regular model set with window $B_n\setminus
B_{n-|z|}$.
Due to Lemma \ref{modsum}, the sum is bounded by the volume of the window,
which is a shell of thickness $|z|$ of the ball of radius $n$.
Thus, the last expression vanishes like $1/n$ as $n\to\infty$.
The limit $\eta(z)$ is a positive definite function, since it is the
limit of the positive definite functions $\eta_n(z)$.
\end{proof}

\begin{theorem}[Autocorrelation]\label{th:dauto}
Let $f:\mathbb R^m \to \mathbb C$ be continuous, 
and $|x|^{m+1+\alpha}|f(x)|\le C$ for some constants $C>0$ and $\alpha>0$.
The weighted dense Dirac comb $\omega$ has the unique autocorrelation
\begin{displaymath}
\gamma_\omega = \sum_{z \in L} \eta(z) \delta_z, \qquad  
\eta(z) = \frac{1}{|\det ({\widetilde L})|} \int_{\mathbb R^m} f(u)
\overline{f(u-z^\star)} \, du,
\end{displaymath}
where $\gamma_\omega$ is a translation bounded, positive definite pure point 
measure.
\end{theorem}

\begin{proof}
Consider the regular model set autocorrelation measures
\begin{displaymath}
\gamma_{\omega,s}=\sum_{z\in \Delta_s}\eta_s(z)\delta_z, \qquad
\eta_s(z) = \frac{1}{|\det ({\widetilde L})|} 
\int_{B_s\cap (B_s+z^\star)} (\chi_s\cdot f)(u) \overline{(\chi_s\cdot
  f)(u-z^\star)}\, du,
\end{displaymath}
where $B_s$ denotes the ball of radius $s\in\mathbb N$, and 
$\Delta_s=\Lambda(B_s)-\Lambda(B_s)$.
Note that $\Delta_s\subset\Delta_{s+1}\subset L$ and set $\eta_s(z)=0$
if $z\notin \Delta_s$.
Consider the associated finite autocorrelation measures
\begin{displaymath}
\gamma_{\omega,s}^{(n)}=\sum_{z\in \Delta_s}\eta_{n,s}(z)\delta_z, \qquad
\eta_{n,s}(z) = \frac{1}{\mbox{vol}(B_n)} 
\sum_{\genfrac{}{}{0pt}{1}{x,y\in \Lambda(B_s) \cap B_n}{x-y=z}}
(\chi_s\cdot f)(x^\star) (\chi_s\cdot f)(y^\star).
\end{displaymath}
Since these measures arise from regular model sets, we have
$\gamma_{\omega,s}^{(n)}\to \gamma_{\omega,s}$ vaguely.
The measures $\gamma_{\omega,s}$ converge to $\gamma_\omega$ vaguely,
as follows from Lebesgue's dominated convergence theorem:
Note that $\eta_s(z)\to\eta(z)$ and 
$|\eta_s(z)|<|\eta|(z):= \frac{1}{|\det ({\widetilde L})|} 
\int_{\mathbb R^m} |f(u)|\, |f(u-z^\star)| du.$
Note that $\gamma_{|\omega|}^{} := \sum_{z\in L} |\eta|(z) \delta_z$ defines a
measure:
For $\gamma_{|\omega|}^{}$ and compact $K$, we have the estimate
\begin{displaymath}
\begin{split}
\gamma_{|\omega|}^{}&(K)
\le \sum_{z\in L\cap W}|\eta|(z)
= \sum_{z^\star\in(L\cap W)^\star} \frac{1}{|\det ({\widetilde L})|}
\int_{\mathbb R^m}|f(u)| \, |f(u-z^\star)|du \\
&=\frac{1}{|\det ({\widetilde L})|}
\int_{\mathbb R^m}|f(u)|
\left( \sum_{z^\star\in(L\cap W)^\star} |f(u-z^\star)|\right) \, du < \infty,
\end{split}
\end{displaymath}
where $W=\bigcup_{i=1}^n W_i$ is a covering of $K$ with a finite number of
translated unit balls $W_i$.
In the last equation, we exchanged summation and integration, which is
justified by Lebesgue's dominated convergence theorem, since the terms in
brackets are bounded uniformly in $u$ due to Lemma \ref{modsum}. 
For $g$ continuous with support on $K$, we then have
\begin{displaymath}
\lim_{s\to\infty} \gamma_{\omega,s}(g) 
= \lim_{s\to\infty} \sum_{z\in L \cap K} \eta_s(z) g(z) 
= \sum_{z\in L \cap K}  \lim_{s\to\infty} \eta_s(z) g(z)
= \sum_{z\in L \cap K}  \eta(z) g(z)
= \gamma_\omega^{}(g).
\end{displaymath}
We will show below that $\gamma_{\omega,s}^{(n)}\to \gamma_{\omega}^{(n)}$
vaguely, uniformly in $n$.
The above results may then be used with a $3\varepsilon$ argument as in Theorem
\ref{Weyl2} to show that $\gamma_{\omega}^{(n)}\to \gamma_{\omega}$.
Since $\omega$ is translation bounded, we further conclude that all finite
volume approximations $\gamma_\omega^{(n)}$ are uniformly translation
bounded (Hof (1995), Prop. 2.2),
hence the limit is translation bounded.
As it coincides with the pointwise limit, it is unique.
The measures $\gamma_\omega^{(n)}$ are positive definite by construction.
Since the positive definite measures are closed in the vague topology,
$\gamma_\omega^{}$ is a positive definite measure.
The explicit form of the vague limit shows that $\gamma_\omega^{}$
is pure point.

To show the uniform convergence of $\gamma_{\omega,s}^{(n)}$, consider for a
compact set $K\subset\mathbb R^d$ the estimate
\begin{displaymath}
\begin{split}
|\gamma_{\omega}^{(n)}(K)&-\gamma_{\omega,s}^{(n)}(K)| =\\
\frac{1}{\mbox{vol}(B_n)}&
\left| 
\sum_{z\in L\cap K} \! \sum_{\genfrac{}{}{0pt}{1}{x\in L \cap B_n}{x-z\in L
    \cap B_n}} \!\!\! f(x^\star) \overline{f(x^\star-z^\star)}
-
\sum_{z\in L\cap K} \! \sum_{\genfrac{}{}{0pt}{1}{x\in \Lambda(B_s) \cap
    B_n}{x-z\in \Lambda(B_s) \cap B_n}} \!\!\! (\chi_s\cdot f)(x^\star)
\overline{(\chi_s\cdot f)(x^\star-z^\star)} 
\right|\\
\le \frac{1}{\mbox{vol}(B_n)}&
\left| 
\sum_{z\in L\cap K} 
\left( \sum_{x\in L \cap B_n} f(x^\star) \overline{f(x^\star-z^\star)}
- \!\!\!
\sum_{x\in \Lambda(B_s) \cap B_n} (\chi_s\cdot f)(x^\star)
\overline{(\chi_s\cdot f)(x^\star-z^\star)} 
\right)
\right|\\
+ \frac{1}{\mbox{vol}(B_n)}&
\left| 
\sum_{z\in L\cap K} 
\left( \sum_{\genfrac{}{}{0pt}{1}{x\in L \cap B_n}{x-z\notin L
    \cap B_n}} \! f(x^\star) \overline{f(x^\star-z^\star)}
-\!\!\!
\sum_{\genfrac{}{}{0pt}{1}{x\in \Lambda(B_s) \cap
    B_n}{x-z\notin \Lambda(B_s) \cap B_n}} \! (\chi_s\cdot f)(x^\star)
\overline{(\chi_s\cdot f)(x^\star-z^\star)} 
\right)
\right|.
\end{split}
\end{displaymath}
Both terms in the last inequality may be estimated by
\begin{displaymath}
\frac{1}{ \mbox{vol}(B_n)}\sum_{z\in L\cap K}
\left(
\sum_{x^\star\in (L\cap B_n)^\star \setminus B_s} |f(x^\star)| \,
|f(x^\star-z^\star)| +
\sum_{\genfrac{}{}{0pt}{1}{x^\star\in (L \cap B_n)^\star}{s-1\le|x^\star|<s}}
|f(x^\star)| \, |f(x^\star-z^\star)|
\right)
\end{displaymath}
by the same argument as in the proof of Theorem \ref{Weyl2}.
We may now interchange the summations and use the fact that
the sum over $z$ is bounded uniformly in $x^\star$ due to Lemma \ref{modsum}.
The remaining term is of the same form as that in Theorem \ref{Weyl2}.
We conclude that it approaches zero as $s\to\infty$ uniformly in $n$.
Thus $\gamma_{\omega,s}^{(n)}\to \gamma_{\omega}^{(n)}$ vaguely and uniformly
in $n$, which concludes the proof.
\end{proof}

It now follows from the theorem of Bochner-Schwartz (Reed and
Simon~(1980), p.~331)
that the Fourier transform of the autocorrelation is a positive, translation
bounded measure.
An explicit expression for the discrete part
$\left(\widehat{\gamma_{\omega}}\right)_{pp}$ of
$\widehat{\gamma_{\omega}}$, which is given in the following theorem,
can be deduced from Hof (1995), Thm 3.4.
The theorem states that the Fourier transform of the
autocorrelation measure has no continuous component.

\begin{theorem}[Diffraction formula]\label{th:ddiff}
Let $f:\mathbb R^m \to \mathbb C$ be continuous
and, for some constants $C>0$ and $\alpha>0$, $|x|^{m+1+\alpha}|f(x)|\le C$.
The Fourier transform $\widehat{\gamma_{\omega}}$ of the autocorrelation
$\gamma_\omega^{}$ of the weighted dense Dirac comb $\omega$ is a
positive, translation bounded pure point measure and explicitly given by
\begin{displaymath}
\widehat{\gamma_{\omega}} =  \sum_{k\in L^*} |c(k)|^2 \delta_k
=\frac{1}{|\det ({\widetilde L})|^2} \sum_{k\in L^*} |\hat
f(-k^\star)|^2 \delta_k,
\end{displaymath}
where the $\mathbb Z$-module $L^*$ is the projection of the dual lattice.
\end{theorem}

\begin{proof}
We showed in the proof of Theorem \ref{th:dauto} that the
autocorrelation measures $\gamma_{\omega,s}$ converge vaguely to
$\gamma_{\omega}$.
Since the Fourier transform is continuous in the vague topology, the
Fourier transforms of the autocorrelation measures
$\widehat{\gamma_{\omega,s}}$ converge vaguely to
$\widehat{\gamma_{\omega}}$.
We will show that the vague limit leads to the above expression.

Take a compact set $K\subset \mathbb R^d$ and a covering $W=\bigcup_{i=1}^n
W_i$ of $K$ with a finite number of translated unit balls $W_i$.
Since $(L^*\cap W)^\star$ is, by duality, a Delone set, we may order the
numbers $k\in L^*\cap K$ to obtain a sequence $(k_i)_{i\in\mathbb N}$ with
$|k^\star_j|\ge |k^\star_i|$ for $j>i$.
Moreover, since $c'_s(k)\to c(k)$ (in the notation of the proof of Theorem
\ref{th:dFB}), we may choose a subsequence $(s_j)_{j\in\mathbb N}$ such that
$\left| |c'_{s_j}(k_i)|^2 - |c(k_i)|^2\right|<2^{-i}$ on $L^*\cap K$ for all
$j\in\mathbb N$.
For explicity of notation, we will suppress the index $j$ in the following.
Fix $\varepsilon>0$.
We have
\begin{displaymath}
\sum_{i=m+1}^\infty |c'_s(k_i)|^2 + \sum_{i=m+1}^\infty |c(k_i)|^2 \le
\sum_{i=m+1}^\infty 2^{-i} + 2 \sum_{i=m+1}^\infty |c(k_i)|^2 < \varepsilon
\end{displaymath}
for $m>m_0(\varepsilon)$.
The last sum can be made arbitrarily small, since it arises from the
discrete part of the measure $\widehat{\gamma_\omega^{}}$, see
Hof (1995), Thm 3.4.
Moreover, we have
$\sum_{i=1}^m \left( |c'_s(k_i)|^2-|c(k_i)|^2\right)<\varepsilon$
for $s>s_0(\varepsilon,m)$, since $|c'_s(k)|^2\to |c(k)|^2$ uniformly in $k$.
Fix now $m>m_0(\varepsilon)$.
For $s>s_0(\varepsilon,m)$, we have
\begin{displaymath}
\left| \sum_{i=1}^\infty |c'_s(k_i)|^2-\sum_{i=1}^\infty |c(k_i)|^2\right| \le
\! \! \sum_{i=m+1}^\infty |c'_s(k_i)|^2+\left|\sum_{i=1}^m \left(
    |c'_s(k_i)|^2-|c(k_i)|^2\right)\right|+\sum_{i=m+1}^\infty |c(k_i)|^2 \le
2\varepsilon.
\end{displaymath}
For $g$ continuous and compactly supported on $K$, this implies 
\begin{displaymath}
 \widehat{\gamma_\omega^{}}(g)
=\lim_{s\to\infty} \widehat{\gamma_{\omega,s}}(g) 
 = \lim_{s\to\infty} \sum_{k\in L^* \cap K} |c'_s(k)|^2 g(k)
= \sum_{k\in L^* \cap K}  |c(k)|^2 g(k).
\end{displaymath}
Since $K$ was an arbitrary compact set, our claim follows.
\end{proof}

The above results lead to a generalized Poisson summation formula.
Recall that, for a lattice $L\subset\mathbb R^d$, the Poisson summation
formula is (C\'ordoba (1989))
\begin{equation}
\left( \sum_{x\in L} \delta_x \right)^{\widehat{}} = 
\frac{1}{|\det (L)|} \sum_{k\in L^*} \delta_k,
\end{equation}
where $|\det (L)|$ is the volume of the fundamental domain of the
lattice $L$, and the sum ranges over all points of the dual lattice $L^*$.
For regular model sets $\Lambda(W)\subset\mathbb R^d$, a generalized Poisson
formula is given by
\begin{equation}
\left( \sum_{z\in \Delta} \eta(z) \delta_z \right)^{\widehat{}} = \sum_{k\in
  L^*} |c_W(k)|^2 \delta_k,
\end{equation}
as is readily inferred from Theorem \ref{th:diff}.
This identity is not symmetric because the first sum ranges over
the uniformly discrete set $\Delta=\Lambda(W)-\Lambda(W)$, whereas the second
sum ranges over the projection of the dual lattice $L^*$, which is a dense set.
For weighted Dirac combs defined on a dense $\mathbb Z$-module
$L\subset\mathbb R^d$, we obtain from Theorem \ref{th:ddiff} the
formula
\begin{equation}
\left( \sum_{z\in L} \eta(z) \delta_z \right)^{\widehat{}} = 
\sum_{k\in L^*} |c(k)|^2 \delta_k,
\end{equation}
where the sums on both sides are defined on a dense set.
Within this symmetric setup, it may be easier to investigate which constraints
on an underlying point set an identity of the above type imposes
(Lagarias (2000), Problem 4.1).

\section{Applications to random tilings}

In this section, we explain how the above results can be used to infer
diffraction properties of random tiling ensembles.

A {\it tiling} of $\mathbb R^d$ is a face-to-face space filling with tiles from
a finite set of $\mathbb R^d$-polytopes called {\it prototiles}, without any
gaps and overlaps (Richard et al.~(1998)).
There might be a number of additional packing rules specifying the allowed 
configurations.
Associated with a tiling is the set $\Lambda$ of all vertex positions of the
tiling.
Take a ball $B_n\subset\mathbb R^d$ of radius $n$ and count the number of
different patches of $B_n$, where we identify patches which are equal up to a
translation.
If the number of allowed patches increases exponentially with the volume of
the ball, we call the set of all tilings a {\it random tiling} ensemble.

The usual ideal quasiperiodic tilings like the Penrose tiling, the
Ammann-Beenker tiling, and others, have the special property that the
vertex positions of its tiles form a (regular) model set.
Essentially, the set of all tilings is given by the collection of the original
model set together with model sets of arbitrarily translated window. 
For tilings of $\mathbb R^d$ whose vertex positions constitute a Delone set and
are described by a primitive substitution, it can be shown that the number of
allowed patches of radius $n$ grows asymptotically proportional to
$n^d$ (Lenz (2002)).

Relaxation of the packing rules for these tilings usually results in a random 
tiling ensemble with strictly positive entropy (Henley (1999)).
Since entropy is an indication of disorder, one should expect a non-vanishing
continuous component in the diffraction spectrum in addition to a discrete
part.
By construction, the set of vertex positions of each random tiling can 
be mapped into internal space via the star-map, resulting in a distribution, 
which may be supported on an unbounded domain in internal space.
This is different from the distribution of a model set $\Lambda(W)$, 
which is the characteristic function $1_W$ of the window.

A natural object to consider is the {\it averaged distribution}, where we take
the average over all random tilings.
Here, we adopt the normalization that $0\in\Lambda$ for all random tilings 
$\Lambda$.
The averaged distribution defines a weighted point set of $\mathbb R^d$ which
is supported on a countable, dense subset of $\mathbb R^d$.

Henley (1999)
gives non-rigorous arguments that the averaged 
distribution (for infinite tilings) exists in dimensions $d>2$.
In dimensions $d\le 2$, the width of the averaged distribution for finite 
patches on balls should diverge with the radius $n$ of the ball.
In $d=1$, the divergence is with the square root of the system size, in $d=2$,
the divergence is logarithmic.  
He concludes that in dimension $d\le 2$ the diffraction spectrum of a random
tiling should have a trivial discrete part.

In $d=1$, his arguments can be made rigorous (Baake et al.~(2002)).
Consider, for example, the Fibonacci model set.
Other tilings may be treated similarly.
Let $\mathbb Z[\tau]=\mathbb Z + \mathbb Z \tau\subset\mathbb R$, where
$\tau=(1+\sqrt{5})/2$ is the golden mean.
The ring $\mathbb Z[\tau]$ is the ring of integers of the quadratic field 
$\mathbb Q(\tau)=\mathbb Q(\sqrt{5})$.
Let $()^\star$ denote the automorphism of $\mathbb Q(\tau)$ that maps
$\sqrt{5}\mapsto-\sqrt{5}$.
The set $\widetilde L=\{(x,x^\star)\,|\, x\in\mathbb Z[\tau]\}$
is a lattice in $\mathbb R\times\mathbb R$.
The Fibonacci model set is the set $\Lambda(W)=\{x\in\mathbb Z[\tau]\,|\,
x^\star \in (-1,\tau-1]\}$.
Consecutive points of the model set have distances $1$ or $\tau$.
If we regard the point positions as left endpoints of half-open intervals 
$u,v$ of length $1$ and $\tau$, we get the {\it Fibonacci tiling}.
A {\it Fibonacci random tiling} is an arbitrary sequence of intervals $u,v$
such that the frequency of $u$ is a.s. equal to $1/\tau$.
The averaged distribution for a patch of size $N$ is, to leading order in $N$, 
given by (H\"offe~(2001), Baake et al.~(2002))
\begin{equation}\label{form:asympt}
\rho(x^\star) = \sqrt{\frac{\tau}{2N}} \, f\left( x^\star
\sqrt{\frac{\tau}{2N}} \right),
\qquad
f(z) = 2 \left( \frac{e^{-z^2}}{\sqrt{\pi}}-|z|\,\mbox{erfc}(|z|)\right),
\end{equation}
where $\mbox{erfc}(x)=\frac{2}{\sqrt{\pi}}\int_x^\infty e^{-t^2}\, dt$ denotes
the complementary error function.
The distribution width therefore grows with the system size as
$\sqrt{N}$.
Together with the above analysis, this may lead to a trivial Bragg
peak at the origin as $N\to\infty$.
This corresponds to the behavior of the Fibonacci random tiling, whose
diffraction spectrum has been computed explicitly (Baake and H\"offe
(2000)).
It consists of an absolutely continuous component in addition to a
trivial Bragg peak at the origin.
It would be interesting to compute the continuous component of the averaged 
structure, which may be given by large $N$ corrections to the asymptotic 
behavior in equation (\ref{form:asympt}).

For $d>1$, there are no rigorous results about the averaged
distribution of quasicrystalline random tilings, which is due to more
restrictive matching rules for the prototiles.  (In $d=1$, the only
matching rule is the face-to-face condition, resulting in Bernoulli
ensembles, which are easy to analyze.)  There are however
numerical investigations for a number of quasicrystallographic random
tilings in $d=2$ and $d=3$.  

In $d=2$, the averaged distribution of the Ammann-Beenker random
tiling appears to be of Gaussian type, with a distribution width diverging
logarithmically with the system size (H\"offe (2001)).  Thus, the
situation is similar to the $d=1$ case discussed above.  A numerical
analysis of the diffraction measure indicates a trivial Bragg peak at
the origin, together with a singular continuous component (H\"offe~(2001)).
This behaviour is believed to be generic in $d=2$ (Henley~(1999)).

In $d\ge 3$, the averaged
distribution is predicted to exist (Henley (1999)).  Our result then
implies that the averaged structure is pure point diffractive, if the
distribution function is well-behaved.  However, the diffraction
picture of a typical random tiling in $d\ge3$ is expected to display
an diffuse background in addition to Bragg peaks (Henley (1999)).
The averaging on the level of the Dirac comb thus has the effect of
extinguishing continuous components of the diffraction spectrum.

To conclude, it is necessary to prove Henley's statements about
the existence of an averaged distribution with finite width in $d\ge3$.
Furthermore, it would be interesting to investigate the relation between 
diffraction properties of the averaged structure and the random tiling, 
in particular whether the discrete parts of both structures coincide and 
whether the continuous parts coincide in $d\le2$. 
Concerning the results of this paper, a natural question is how they
can be extended beyond the Euclidean case towards the setup of locally
compact Abelian groups.
(The proofs of the pure pointedness of the diffraction spectrum used
the fact that $\mathbb R^n$ is isomorphic to its dual.) 
Another aspect concerns the connection to the theory of almost periodic
measures (Gil de Lamadrid and Argabright (1990)),
which may also be used for a characterization of diffraction spectra
with continuous components.

\section*{Acknowledgments}

The author thanks Michael Baake and Moritz H\"offe for discussions,
and Michael Baake and Bernd Sing for a critical reading of 
the manuscript.
The author is grateful to the Erwin Schr \"odinger International
Institute for Mathematical Physics in Vienna for supporting a stay in
December 2002, where the manuscript was completed.
Financial support from the German Research Council (DFG) is gratefully
acknowledged.


\begin{thebibliography}{99}

\bibitem{BH00} 
Baake, M. and H\"offe, M.,
{\it Diffraction of random tilings: Some rigorous results},
J.~Stat.~Phys. {\bf 99}, 219-261 (2000);
math-ph/9904005.

\bibitem{BM98}
Baake, M. and Moody, R.V.,
{\it Diffractive point sets with entropy},
J.~Phys.~A: Math.~Gen. {\bf 31}, 9023-9039 (1998);
math-ph/9809002.

\bibitem{BM00}
Baake, M. and Moody, R.V.,
{\it Self-similar measures for quasicrystals}, in:
{\it Directions in Mathematical Quasicrystals},
eds. Baake, M. and Moody, R.V.,
CRM Monograph Series, vol. 13, AMS, Providence, RI (2000),
pp. 1-42; math.MG/0008063.

\bibitem{BM02}
Baake, M. and Moody, R.V.,
{\it Weighted Dirac combs with pure point diffraction},
preprint (2002); math.MG/0203030.

\bibitem{BMP00}
Baake, M., Moody, R.V. and Pleasants, P.A.B.,
{\it Diffraction from visible lattice points and $k$th 
power free integers},
Discr.~Math.~{\bf 221}, 3-42 (2000);
math.MG/9906132.

\bibitem{BMRS02}
Baake, M., Moody, R.V., Richard, C. and Sing, B.,
{\it Which distributions of matter diffract? -- Some answers}, in:
{\it Quasicrystals: Structure and Physical Properties},
ed. Trebin, H.-R., Wiley-VCH, Berlin (2003), pp. 188-207;
math-ph/0301019.

\bibitem{BD00}
Bernuau, G. and Duneau, M.,
{\it Fourier analysis of deformed model sets}, in:
{\it Directions in Mathematical Quasicrystals},
eds. M. Baake and R.V. Moody,
CRM Monograph Series, vol. 13, AMS, Providence, RI (2000),
pp. 43-60.

\bibitem{BT86}
Bombieri, E. and Taylor, J.E.,
{\it Which distributions of matter diffract? An initial
investigation},
J.~Phys.~{\bf 47}, Colloq. C3, Suppl. au No.7, 19-28 (1986).

\bibitem{C89}
C\'ordoba, A.,
{\it Dirac combs},
Lett.~Math.~Phys. {\bf 17} 191-196 (1989).

\bibitem{D93}
Dworkin, S.,
{\it Spectral theory and X-ray diffraction},
J.~Math.~Phys. {\bf 34}, 2965-2967 (1993).

\bibitem{GA90}
Gil de Lamadrid, J. and Argabright, L.N.,
{\it Almost Periodic Measures},
Memoirs of the AMS, vol. {\bf 428},
AMS, Providence, RI (1990).

\bibitem{Hen99}
Henley, C.L.,
{\it Random tiling models}, in:
{\it Quasicrystals: The State of the Art}, 
eds: D.~P.~DiVincenzo and P.~J.~Steinhardt, 
Series on Directions in Condensed Matter Physics, vol. 16,
World Scientific, Singapore, 2nd edition (1999), 
pp. 459-560.

\bibitem{HB00}
H\"offe, M. and Baake, M.,
{\it Surprises in diffuse scattering},
Z.~Kristallogr. {\bf 215}, 441-444 (2000);
math-ph/0004022.

\bibitem{H01}
H\"offe, M.,
{\it Diffraktionstheorie stochastischer Parkettierungen},
Dissertation, Universit\"at T\"ubingen, Shaker, Aachen (2001).

\bibitem{H95}
Hof, A.,
{\it On diffraction by aperiodic structures}, 
Commun. Math. Phys. {\bf 169}, 25-43 (1995).

\bibitem{H98}
Hof, A.,
{\it Uniform distribution and the projection method}, in: 
{\it Quasicrystals and Discrete Geometry}, 
ed. J. Patera, 
Fields Institute Monographs, vol. 10, AMS, Providence, RI (1998), 
pp. 201-206.

\bibitem{KN74}
Kuipers, L. and Niederreiter, H.,
{\it Uniform Distribution of Sequences},
Wiley, New York, 1974.

\bibitem{K01}
K\"ulske, C.,
{\it Universal bounds on the selfaveraging of random diffraction measures},
WIAS-preprint {\bf 676} (2001);
math-ph/0109005.

\bibitem{L00}
Lagarias, J.C.,
{\it Mathematical quasicrystals and the problem of diffraction}, in:
{\it Directions in Mathematical Quasicrystals},
eds. M. Baake and R.V. Moody,
CRM Monograph Series, vol. 13, AMS, Providence, RI (2000),
pp. 61-93.

\bibitem{L02}
Lenz, D., private communication (2002).

\bibitem{M97}
Moody, R.V.,
{\it Meyer sets and their duals}, in:
{\it The Mathematics of Long-Range Aperiodic Order},
ed. R.V. Moody,
NATO ASI Series C 489,
Kluwer, Dordrecht (1997),
pp. 403-441.

\bibitem{RS80}
Reed, M. and Simon, B.,
{\it Methods of Modern Mathematical Physics. I: Functional Analysis},
2nd ed., Academic Press, San Diego, CA (1980).

\bibitem{RHHB98}
Richard, C., H\"offe, M., Hermisson, J., and Baake, M.,
{\it Random tilings: Concepts and examples},
J.~Phys.~A:~Math.~Gen. {\bf 31}, 6385-6408 (1998);
cond-mat/9712267.

\bibitem{R99}
Richard, C.,
{\it An alternative view on quasicrystalline random tilings},
J.~Phys.~A: Math.~Gen. {\bf 32}, 8823-8829 (1999);
cond-mat/9907262.

\bibitem{Sch98}
Schlottmann, M.,
{\it Cut-and-project sets in locally compact Abelian groups}, in: 
{\it Quasicrystals and Discrete Geometry}, 
ed. J. Patera, 
Fields Institute Monographs, vol. 10, AMS, Providence, RI (1998), 
pp. 247-264.

\bibitem{Sch00}
Schlottmann, M.,
{\it Generalized model sets and dynamical systems}, in:
{\it Directions in Mathematical Quasicrystals},
eds. M. Baake and R.V. Moody,
CRM Monograph Series, vol. 13, AMS, Providence, RI (2000),
pp. 143-159.

\end{thebibliography}
\end{document}